\documentclass{ws-procs975x65}

\def\be{\begin{equation}}
\def\ee{\end{equation}}
\def\bea{\begin{eqnarray}}
\def\eea{\end{eqnarray}}

\begin{document}

\title{Cosmological implications of Geometrothermodynamics}

\author{Orlando Luongo${}^{1,2,3}$ and Hernando Quevedo${}^{3,4,5}$}
\address{${}^{1}$Dip. di Fisica, Universit\`a di Napoli "Federico
II", Napoli, Italy;\\
${}^{2}$INFN, Universit\`a di Napoli "Federico
II", Napoli, Italy;\\
${}^{3}$Inst. de Ciencias  Nucleares, Universidad  Nacional Aut\'onoma de
M\'exico,  M\'exico DF, Mexico\\
${}^{4}$Dip. di Fisica and ICRA, Universit\`a di Roma "La Sapienza", Roma, Italy.\\
${}^{5}$Inst. de Cosmologia, Relatividade e Astrof\'\i sica, ICRA - CBPF, Rio de Janeiro, Brazil.
}

\begin{abstract}
We use the formalism of Geometrothermodynamics to derive a series of fundamental equations for thermodynamic systems. It is shown that
all these fundamental equations can be used in the context of relativistic cosmology to derive diverse scenarios  which include the standard
cosmological model, a unified model for dark energy and dark matter, and an effective inflationary model.
\end{abstract}

\section{Introduction}

In the context of modern cosmology, the issue of determining the energy content which drives the cosmological acceleration is still now object of debate \cite{copeland}. The fluid responsible for the observed cosmic speed up is commonly called dark energy. The adjective ``dark" is due to the lack on knowledge of its physical properties, since the fluid shows repulsive effects able to counterbalance the attraction of gravity. Here, we present several cosmological models, and propose a way to describe the observed late time acceleration by using results obtained from the so-called Geometrothermodynamics (GTD) \cite{ddt}. In this way, one infers from GTD the physical meaning of dark energy as a thermodynamic effect.

GTD is represented in a $(2n+1)-$dimensional contact differential manifold ${\cal T}$, whose coordinates are $Z^A =\{\Phi, E^a, I ^a\}$ with $a=1,...,n$, and $A=0,1,...,2n$. Here, $\Phi$ represents the thermodynamic potential, while $E^a$ correspond to the extensive variables and $I^a$ to the intensive variables. The contact structure of ${\cal T}$ is determined by the fundamental one-form $\Theta=d\Phi - \delta_{ab} I^adE^b$, which is defined modulo an arbitrary conformal factor. One of the basic ingredients of GTD is its invariance with respect to Legendre transformations. They are defined as transformations of the form $\{Z^A\}\longrightarrow \{\widetilde{Z}^A\}=\{\tilde \Phi, \tilde E ^a, \tilde I ^ a\}$ with $ \Phi = \tilde \Phi - \delta_{kl} \tilde E ^k \tilde I ^l ,\ E^i = - \tilde I ^ {i}, \  E^j = \tilde E ^j,\   I^{i} = \tilde E ^ i , \ I^j = \tilde I ^j$, where $i\cup j$ is any disjoint decomposition of the set of indices $\{1,...,n\}$, and $k,l= 1,...,i$. The most general metric, we have found so far, which is invariant with respect to partial and total Legendre transformations is given by
\begin{equation}
\label{kjd}
G=\left(d\Phi - I_a dE^a\right)^2 + \Lambda\, (E_a I_a)^{2k+1} d E^a d I^a\,, \quad E_a =\delta_{ab}E^b\ ,\ I_a =\delta_{ab} I ^b\ ,
\end{equation}
where $\Lambda$ is an arbitrary Legendre invariant function of $Z^A$, and $k$ is an integer.

\section{Cosmology in the equilibrium manifold}

The equilibrium manifold ${\cal E}$ is an $n-$dimensional submanifold ${\cal E} \subset {\cal T}$, defined by the smooth embedding map $\varphi:{\cal E} \rightarrow {\cal T}$, such that the condition $\varphi^*(\Theta)=0$ holds, which is equivalent to the first law of thermodynamics, $d\Phi = I_a dE^a$, if $E^a$ are considered as the coordinates of ${\cal E}$. Moreover, ${\cal E}$ is a Riemannian manifold endowed with the induced metric $g=\varphi^*(G)$, which in the case of (\ref{kjd})
yields
\begin{equation}
g=\Lambda \left(E_a\Phi_a\right)^{2k+1} \delta^{ab}\Phi_{bc} dE^a dE^c\ ,\quad
\Phi_a = \frac{\partial \Phi}{\partial E^a}\ ,\quad \Phi_{bc} = \frac{\partial^2 \Phi}{\partial E^b\partial E^c}\ .
\end{equation}
The thermodynamic systems ``live" in ${\cal E}$ and the objective of GTD is to relate the geometric properties of ${\cal E}$ with the thermodynamic properties of the system, according to the following scheme:

1. the curvature of ${\cal E}$ is a measure of the thermodynamic interaction;

2. curvature singularities of ${\cal E}$ represent phase transitions;

3. thermodynamic geodesics of ${\cal E}$ represent quasi-static processes.

\noindent If, in addition, we demand that the equilibrium manifold be an extremal surface embedded in ${\cal T}$, i.e., that
$\frac{\delta}{\delta Z^A} \int_{\cal E} \sqrt{{\rm det}(g)} d^n E=0$, then the solutions of the corresponding field equations
$\Box Z^A=0$ can be interpreted as fundamental equations $\Phi=\Phi(E^a)$. Consider, for instance, the case of systems with two
thermodynamic degrees of freedom $(n=2)$ in the entropy representation $\Phi=S$. Then, one can show that the conformal factor $\Lambda$, which
enters the metric (\ref{kjd}), can be chosen such that the fundamental equations \cite{vqs10}
\begin{equation}
S_1 =c_1\ln U + c_2 \ln V  ,\ S_2 = S_0\ln (U^\alpha + c U^\beta) , \ S_3 = c_1 \ln \left( U + \frac{a}{V}\right) + c_2 \ln (V-b),\nonumber
\end{equation}
represent extremal surfaces of the phase manifold, where $a$, $b$, $c_1$, $c_2$, $\alpha$, $\beta$ and $S_0$ are arbitrary real constants, and $U$ and $V$ are the internal energy and volume, respectively. The corresponding metrics and curvatures for the equilibrium manifold can be calculated in a straightforward manner. It follows that $S_1$ determines a flat space, indicating that the system has no thermodynamic interaction, $S_2$ determines a curved space with no singularities, and, finally, $S_3$ corresponds to a space with curvature singularities, i.e., a thermodynamic system with phase transitions. It turns out that all the above fundamental equations describe the thermodynamics of different cosmological models. In fact, by using the equilibrium conditions $I_a=\frac{\partial \Phi}{\partial E^a}$, from the expression for $S_1$ we find the barotropic equation of state (EoS) $P=\frac{c_2}{c_1}\frac{U}{V}$ that can be used as the EoS for a Friedman-Robertson-Walker model in the context of relativistic cosmology. Then, for $\frac{c_2}{c_1} =\frac{1}{3}$ we obtain a radiation dominated universe, for $c_2=0$ a matter dominated universe, and for $\frac{c_2}{c_1} =-1$ an accelerated universe dominated by dark energy. It is interesting to notice that this simple result opens an alternative way to investigate the nature of dark energy. Indeed, since the heat capacity is $C_V=\frac{\partial U}{\partial T}=c_1$, from the condition $\frac{c_2}{c_1} =-1$, we obtain
\begin{equation}
U=C_VT \quad {\rm and}\quad  PV=-C_V T\ .
\end{equation}
If we assume $C_V>0$ with positive volume and temperature, it follows that the pressure must be negative. This is the standard interpretation of dark energy. However, a second possibility is to consider $P>0$, implying that $C_V<0$. Furthermore, it is known that systems with negative heat capacity are in general not extensive \cite{bell99}. It then follows that dark energy can be considered as a non-extensive thermodynamic system with positive pressure and negative heat capacity. On the other hand, a statistical analysis shows that non-extensive systems with negative heat capacity are due to the presence of long-range interactions \cite{bell99}, which is the dominant interaction at large scales in the Universe,  and are common in fractal systems. The question arises whether dark energy can be considered as a fractal system with a long-range interaction. A more detailed analysis will be necessary to answer this question.

Consider now the fundamental equation $S_2$. The corresponding EoS can be expressed as
\begin{equation}\label{o}
P=c\frac{(1+\beta)V^\beta}{(1+\alpha)U^\alpha}\,,
\end{equation}
and it represents an extension of the generalized Chaplygin gas model. It can be shown that this EoS can be used to generate a unified model for dark energy and dark matter which is in agreement with cosmological observations \cite{abcq12}.
 In the particular case $\alpha=\beta$, the model shows a constant pressure with an evolving density. This case mimics the $\Lambda$CDM model, showing that it is possible to accelerate the Universe by using a unique fluid approach in which the EoS is proportional to the inverse of the total density. This case has been proposed and extensively investigated by Luongo and Quevedo \cite{luongo}.

The fundamental equation $S_3$ is a generalization of the van der Waals gas. It reduces to $S_1$ in the limit $a,b\rightarrow 0$ so that it contains, as special cases, all the fluids of the standard cosmological model. Since the equilibrium manifold for $S_3$ predicts the existence of phase transitions, which are important during the inflationary era of the Universe, we can expect that $S_3$ describes an effective inflationary model. Indeed, one can show that in the region where phase transitions can occur, the parameter $b$ can be associated with the minimum volume of the Universe (Planck volume) and the value of the parameter $a$ can be derived from the physical quantities that characterize inflation. A more detailed analysis of this effective inflationary model will be presented elsewhere.


\begin{thebibliography}{99}

\bibitem{copeland}
E. J. Copeland, M. Sami, S. Tsujikawa, Int. J. Mod. Phys. D, 15,
1753, (2006).

\bibitem{ddt}
H. Quevedo, J. Math. Phys. 48, 013506, (2007).

\bibitem{vqs10}
A. Vazquez, H. Quevedo, and A. Sanchez, J. Geom. Phys.
60, 1942, (2010).

\bibitem{bell99} D. Lynden-Bell, Physica A, 263, 293, (1999).

\bibitem{abcq12} A. Aviles, A. Bastarrachea, L. Campuzano and H. Quevedo, Phys. Rev. D, 86, 063508, (2012).

\bibitem{luongo}
O. Luongo, H. Quevedo, Astrophys. Sp. Sci., 338, 2, 345, (2012).


\end{thebibliography}
\end{document}